\documentclass[useAMS,usenatbib,fleqn]{mn2e}

\usepackage{graphicx,amssymb}
\usepackage{epsf,psfig}
\usepackage{natbib}
\usepackage{amsmath}
\usepackage[unicode]{hyperref}
\hypersetup{pdftitle = The title of my PDF, pdfauthor = My name, pdfsubject= The subject, pdfkeywords = keyword1 keyword2 keyword3}
\hypersetup{colorlinks = true, linkcolor = blue, anchorcolor = red, citecolor = blue, filecolor = red, pagecolor = red, urlcolor = blue}

\title[Escape dynamics in tidally limited star cluster models]
{Comparing the escape dynamics in tidally limited star cluster models}

\author[Euaggelos E. Zotos]{Euaggelos E. Zotos\thanks{E-mail: evzotos@physics.auth.gr} \\
Department of Physics, School of Science, Aristotle University of Thessaloniki,
GR-541 24, Thessaloniki, Greece}

\begin{document}

\date{Accepted 2015 June 9. Received 2015 June 8; in original form 2015 April 22}

\pubyear{2015} \volume{452} \pagerange{193--209}

\setcounter{page}{193}

\maketitle

\label{firstpage}

\begin{abstract}

The aim of this work is to compare the orbital dynamics in three different models describing the properties of a star cluster rotating around its parent galaxy in a circular orbit. In particular, we use the isochrone and the Hernquist potentials to model the spherically symmetric star cluster and we compare our results with the corresponding ones of a previous work in which the Plummer model was applied for the same purpose. Our analysis takes place both in the configuration $(x,y)$ and in the phase $(x,\dot{x})$ space in order to elucidate the escape process as well as the overall orbital properties of the tidally limited star cluster. We restrict our investigation into two dimensions and we conduct a thorough numerical analysis distinguishing between ordered and chaotic orbits as well as between trapped and escaping orbits, considering only unbounded motion for several energy levels above the critical escape energy. It is of particular interest to determine the escape basins towards the two exit channels (near the Lagrangian points $L_1$ and $L_2$) and relate them with the corresponding escape times of the orbits.

\end{abstract}


\section{Introduction}
\label{intro}

We consider the simplest case of a star cluster in a circular orbit around its parent galaxy. Moreover, the star cluster is embedded in the tidal filed of the parent galaxy with linear tidal forces. Usually we use the term ``tidally limited" star cluster which means, roughly speaking, that the tidal field imposes a boundary, outside of which the mass density vanishes, while the star cluster is being captured within the tidal limit. Stars escape from tidally limited star clusters following a two stages process. During the first stage, stars are being moved by certain physical processes into the escaping phase space, while in the second stage they escape through the channels in the open equipotential surface. The time needed for a star to complete stage two is mainly related with the energy of the star. \citet{FH00} derived an expression for the escape times of stars in second stage which have just finished stage one, while \citet{B01} exploited these results in order to address the important issue of dissolution time of star clusters moving in circular orbits around their parent galaxy.

Escaping particles from dynamical systems is a subject to which has been devoted many studies over the years. Especially the issue of escapes in Hamiltonian systems is directly related to the problem of chaotic scattering which has been an active field of research over the last decades and it still remains open \citep[e.g.,][]{BGOB88,BOG89,JS88,CK92,ML02,SASL06,SSL07,SS08,SHSL09,SS10}. The problem of escape is a classical problem in simple Hamiltonian nonlinear systems \citep[e.g.,][]{AVS01,AS03,AVS09,BBS08,BSBS12,Z14,Z15c} as well as in dynamical astronomy \citep[e.g.,][]{HB83,BTS96,BST98,dML00,N04,N05,dAT14,Z12,Z15b,Z15d}.

The chaotic dynamics of a star cluster embedded in the tidal field of a parent galaxy was investigated in \citet{EJSP08} (hereafter Paper I). Conducting a thorough scanning of the available phase space the authors managed to obtain the basins of escape and the respective escape rates of the orbits, revealing that the higher escape times correspond to initial conditions of orbits near the fractal basin boundaries. The investigation was expanded into three dimensions in \citet{Z15a} where we revealed the escape mechanism of three-dimensional orbits in a tidally limited star cluster. Furthermore, \citet{EP14} explored the escape dynamics in the close vicinity of and within the critical area in a two-dimensional barred galaxy potential, identifying the escape basins both in the phase and the configuration space.

In Paper I the authors used the Plummer potential in order to model the properties of the star cluster. However there are also other spherically symmetric potentials which can also be used for the same purpose. For example there is the H\'{e}non's isochrone and the Hernquist potential, where both of them have been previously used to describe the properties of a star cluster \citep[e.g.,][]{BDL12,APF06}. In the present work, we shall use these two potentials in an attempt to compare the orbital dynamics of the star cluster with the corresponding outcomes of Paper I.

The paper is organized as follows: A detailed presentation of the tidal approximation model is included in Section \ref{mod}. In Section \ref{cometh} we describe all the computational techniques we used in order to determine the nature (bounded or unbounded) of the orbits. The following Section contains all the numerical results showing how the value of the total orbital energy influences the orbital content of the models. Our paper ends with Section \ref{disc}, where the conclusions and the discussion of this work are given.

\section{The tidal approximation model}
\label{mod}

Let us briefly recall the properties of the tidal approximation theory. This model allow us to assume that the star cluster moves with constant angular velocity $\omega$ around the center of the parent galaxy. Using this model we can apply epicyclic approximation to calculate linear tidal forces acting on all members of the cluster. Undoubtedly, the most appropriate system of coordinates is an accelerated rotating reference frame \citep{C42} in which both the the star cluster and the galactic center are at rest, while the origin of the coordinates is at the center of the star cluster which is located at the extremum of the effective potential of the galaxy, so that the position of the galactic center is $\left(-R_{\rm g},0,0\right)$, where $R_{\rm g}$ is the radius of the circular orbit. If the motion of the star cluster around its parent galaxy is more general (i.e., following an elliptical orbit), then no integral of motion comparable to the Jacobi integral is known and therefore all the numerical computations become more difficult to be performed.

In order to model the properties of the star cluster we use two types of spherically symmetric potentials:

(a). The H\'{e}non's isochrone potential \citep{H59}
\begin{equation}
\Phi_{\rm I}(x,y,z) = - \frac{G M_{\rm cl}}{r_{\rm I} + \sqrt{r^2 + r_{\rm I}^2}},
\label{pot1}
\end{equation}
where $r^2 = x^2 + y^2 + z^2$, $G$ is the gravitational constant, $M_{\rm cl}$ is the total mass of the star cluster, while $r_{\rm I}$ is the Isochrone radius similar to the Plummer radius.

(b). The Hernquist potential \citep{H90}
\begin{equation}
\Phi_{\rm H}(x,y,z) = - \frac{\ G M_{\rm cl}}{r_{\rm H} + r},
\label{pot2}
\end{equation}
where $r_{\rm H}$ is the Hernquist radius. Potential (\ref{pot2}) is in fact a Dehnen model \citep{D93} with $\gamma = 1$.

The total effective potential
\begin{equation}
\Phi_{\rm eff}(x,y,z) = \Phi_{\rm cl}(x,y,z) + \frac{1}{2}\left(\kappa^2 - 4\omega^2 \right) x^2 + \frac{1}{2}\nu^2 z^2,
\label{eff}
\end{equation}
where $\Phi_{\rm cl}$ is either the isochrone or the Hernquist potential.

A fundamental scale length of the star cluster is the tidal radius \citep{K62} which is defined as
\begin{equation}
r_{\rm t} = \left(\frac{G M(r_{\rm t})}{4\omega^2 - \kappa^2}\right)^{1/3},
\label{rt}
\end{equation}
where $M(r_{\rm t})$ is the mass contained within the tidal radius and usually is smaller than the total mass of the star cluster $M_{\rm cl}$.

We use the dimensionless system of units introduced in Paper I according to which $G = \omega = r_t = 1$. The formula which relates the physical units of time, mass and radius is $G = 1/222.3$ pc$^3$/M$_{\odot}$/Myr$^2$. The characteristic frequencies $\omega$, $\kappa$ and $\nu$ on the other hand, are related to the galactic gravitational potential and in the solar neighborhood can be expressed as functions of the Oort's constants \citep[see e.g.,][]{BT08} as follows
\begin{align}
\omega^2 &= \left(A - B\right)^2, \\
\kappa^2 &= - 4B\left(A - B\right)^2, \\
\nu^2 &= 4\pi G \rho_{\rm g} + 2 \left(A^2 - B^2\right),
\label{freqs}
\end{align}
where $\rho_{\rm g}$ is the local value of the galactic density \citep{HF00}. Exploiting the numerical values of the Oort's constants provided in \citet{FW97}, we obtain the dimensionless quantities $\kappa^2/\omega^2 \simeq 1.8$ and $\nu^2/\omega^2 \simeq 7.6$.

In both models we choose that $M_{\rm cl} = 2.7$, while the mass contained within the tidal radius is $M(r_{\rm t}) = 2.2$, which means that about 18.5\% of the total mass lies beyond the tidal radius. For the value $M(r_{\rm t}) = 2.2$ we obtain form Eq. (\ref{rt}) that the tidal radius is $r_{\rm t} = 1$ since $4\omega^2 - \kappa^2 = 2.2$. The second step is to set the values of the isochrone and the Hernquist radius. For this task we need a formula that relates the mass with the scale radius. We know that in spherical potentials
\begin{equation}
\frac{d\Phi}{dr} = \frac{GM(r)}{r^2},
\end{equation}
so the cumulative mass distribution is
\begin{equation}
M(r) = \frac{r^2}{G}\frac{d\Phi}{dr}.
\label{massd}
\end{equation}
Inserting Eqs. (\ref{pot1}) and (\ref{pot2}) in formula (\ref{massd}) we have
\begin{equation}
M(r) = \frac{M_{\rm cl}r^3}{\sqrt{r^2 + r_{\rm I}^2}\left(r_{\rm I} + \sqrt{r^2 + r_{\rm I}^2}\right)^2},
\label{md1}
\end{equation}
for the isochrone potential and
\begin{equation}
M(r) = \frac{M_{\rm cl}r^2}{\left(r + r_{\rm H}\right)^2}.
\label{md2}
\end{equation}
for the Hernquist potential (see also Eq. (3) in \citet{H90}).
We already know that $M_{\rm cl} = 2.7$, while for $r = r_{\rm t} = 1$ it is $M(r = r_{\rm t}) = 2.2$. Replacing these values in Eqs. (\ref{md1}) and (\ref{md2}) and solving for $r$ while keeping always the positive solution we finally obtain that $r_{\rm I} = 0.10$, while $r_{\rm H} = 0.107823$. These values secure that the two Lagrangian saddle points $L_1$ and $L_2$ are located exactly at $(x,y,z) = (-1, 0, 0)$ and $(x,y,z) = (1, 0, 0)$, respectively.

Taking into account the relationships connecting the tidal forces with the epicyclic frequency $\kappa$ and the vertical frequency $\nu$ the basic equations of motion are
\begin{align}
\ddot{x} &= - \frac{\partial \Phi_{\rm cl}}{\partial x} - \left(\kappa^2 - 4 \omega^2 \right)x + 2 \omega \dot{y},\\
\ddot{y} &= - \frac{\partial \Phi_{\rm cl}}{\partial y} - 2 \omega \dot{x},\\
\ddot{z} &= - \frac{\partial \Phi_{\rm cl}}{\partial z} - \nu^2 z,
\label{eqmot}
\end{align}
where, as usual, the dot indicates derivative with respect to the time. We observe that both centrifugal and Coriolis forces appear in the above equations of motion due to the fact that the coordinate system is rotating.

In addition, the equations of motion admit the following isolating integral of motion
\begin{equation}
C = \frac{1}{2} \left(\dot{x}^2 + \dot{y}^2 + \dot{z}^2 \right) + \Phi_{\rm eff}(x,y,z) = E,
\label{ham}
\end{equation}
where $\dot{x}$, $\dot{y}$ and $\dot{z}$ are the momenta per unit mass, conjugate to $x$, $y$ and $z$, respectively, while $E$ is the numerical value of the Jacobian integral, which is conserved. The numerical value of the effective potential at the two Lagrangian points $\Phi_{\rm eff}(-1,0,0)$ and $\Phi_{\rm eff}(1,0,0)$ yields to a critical Jacobi constant $C_L$, which can be used to define a dimensionless energy parameter as
\begin{equation}
\widehat{C} = \frac{C_L - E}{C_L},
\label{chat}
\end{equation}
where $E < 0$ is some other value of the Jacobian. The dimensionless energy parameter $\widehat{C}$ makes the reference to energy levels more convenient. For $E = C_L$ the equipotential surface encloses the critical volume, while for a Jacobian value $E > C_L$, or in other words $\widehat{C} > 0$, the equipotential surface is open and consequently stars can escape from the cluster. In the case of the isochrone potential $C_{L_{\rm I}} = -3.543290584540038$, while for the Hernquist potential we have $C_{L_{\rm H}} = -3.537211521390788$.

\begin{figure}
\includegraphics[width=\hsize]{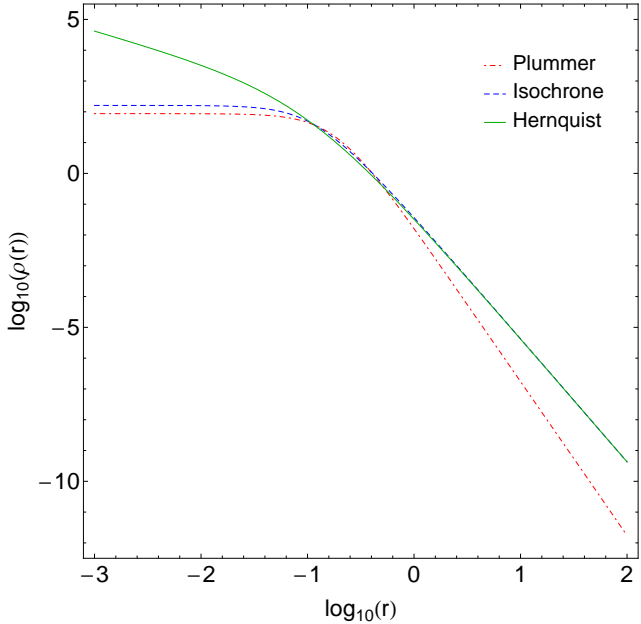}
\caption{Evolution of the mass density of the three spherically symmetric potentials (Plummer, isochrone and Hernquist) as a function of the radius $r$.}
\label{dens}
\end{figure}

It would be very informative to study and compare the mass density derived from the potentials we use to describe the star cluster. Since all potentials are spherically symmetric the mass density can be easily computed using the following version of the Poisson's equation
\begin{equation}
\rho(r) = \frac{1}{4\pi G}\frac{1}{r^2}\frac{d}{dr}\left(r^2 \frac{d\Phi_{\rm cl}}{dr} \right)
\label{den}
\end{equation}
After trivial computations we have
\[
\rho_{\rm P}(r) = \frac{3M_{\rm cl}}{4\pi r_{\rm P}^3}\left(1 + \frac{r^2}{r_{\rm P}^2}\right)^{5/2}
\]
\[
\rho_{\rm I}(r) = \frac{3M_{\rm cl}}{4\pi}\frac{\left(\left(r_{\rm I} + a\right)a^2 - r^2\left(r_{\rm I} + 3a\right)\right)}{\left(r_{\rm I} + a\right)^3 a^3}
\]
\begin{equation}
\rho_{\rm H}(r) = \frac{M_{\rm cl}}{2\pi}\frac{r_{\rm H}}{r}\frac{1}{\left(r + r_{\rm H}\right)^3},
\label{3den}
\end{equation}
where $a = \left(r_{\rm I}^2 + r^2\right)^{1/2}$, while $\rho_{\rm P}(r)$, $\rho_{\rm I}(r)$, $\rho_{\rm H}(r)$ are the densities corresponding to the Plummer, isochrone and Hernquist potential, respectively. In Fig. \ref{dens} we present the evolution of the mass density of the three potentials as a function of the radius $r$. For the Plummer model we take $M_{\rm cl} = 2.7$ and $c_{\rm P} = 0.38$. The value of $c_{\rm P}$ was calculated as the other two scale radii. It is seen that for distances very close to the center $(r < 0.1)$ the density of the Plummer and isochrone potentials remain constant having finite-density core, while on the other hand the density of the Hernquist model increases due to the gentle power-law cusp in the center. For intermediate distances $(0.1 < r < 1)$ the evolution of all three density profiles is almost the same. We also observe that for large radii the asymptotic behavior of the isochrone and the Hernquist mass density profiles is identical varying like $1/r^4$, while the density of the Plummer potential drops more quickly following the power law $1/r^5$. Before closing this section we would like to add that \citet{MLF13} showed that King \citep{K66} and Wilson \citep{W75} density profiles fit globular clusters well. Both King and Wilson models have, in contrast to the Hernquist model, finite densities in the core.

\section{Computational methods}
\label{cometh}

In order to investigate the orbital content of the star cluster, we need to define samples of initial conditions of orbits whose properties (bounded or escaping motion) will be identified. For this purpose, we define for each value of the total orbital energy (all tested energy levels are above the escape energy), dense uniform grids of $1024 \times 1024$ initial conditions regularly distributed in the area allowed by the value of the energy. Our investigation takes place both in the configuration $(x,y)$ and in the phase $(x,\dot{x})$ space for a better understanding of the escape mechanism. Furthermore, the grids of initial conditions of orbits whose properties will be explored are defined as follows: For the configuration $(x,y)$ space we consider orbits with initial conditions $(x_0, y_0)$ with $\dot{y_0} = 0$, while the initial value of $\dot{x_0}$ is always obtained from the Jacobian integral (\ref{ham}) as $\dot{x_0} = \dot{x}(x_0,y_0,\dot{y_0},E) > 0$. Similarly, for the phase $(x,\dot{x})$ space we consider orbits with initial conditions $(x_0, \dot{x_0})$ with $y_0 = 0$, while this time the value of $\dot{y_0}$ is obtained from the Jacobian integral (\ref{ham}).

The equations of motion as well as the variational equations for the initial conditions of all orbits were integrated using a double precision Bulirsch-Stoer \verb!FORTRAN 77! algorithm \citep[e.g.,][]{PTVF92} with a small time step of order of $10^{-2}$, which is sufficient enough for the desired accuracy of our computations (i.e., our results practically do not change by halving the time step). Here we should emphasize, that our previous numerical experience suggests that the Bulirsch-Stoer integrator is both faster and more accurate than a double precision Runge-Kutta-Fehlberg algorithm of order 7 with Cash-Karp coefficients \citep[e.g.,][]{DMCG12}. Throughout all our computations, the Jacobian energy integral (Eq. (\ref{ham})) was conserved better than one part in $10^{-11}$, although for most orbits it was better than one part in $10^{-12}$.

An issue of paramount importance in dynamical systems with escapes is the determination of the position as well as the time at which an orbit escapes from the system. For all energy values smaller than the critical value $C_L$ (escape energy), the 3D equipotential surface is closed. On the other hand, when $E > C_L$ the equipotential surface is open and extend to infinity. The value of the energy itself however, does not furnish a sufficient condition for escape. An open equipotential surface consists of two branches forming channels through which an orbit can escape to infinity. It was proved \citep{C79} that in Hamiltonian systems at every opening there is a highly unstable periodic orbit close to the line of maximum potential which is called a Lyapunov orbit. Such an orbit reaches the Zero Velocity Curve (ZVC), on both sides of the opening and returns along the same path thus, connecting two opposite branches of the ZVC. Lyapunov orbits are very important for the escapes from the system, since if an orbit intersects any one of these orbits with velocity pointing outwards moves always outwards and eventually escapes from the system without any further intersections with the surface of section \citep[e.g.,][]{C90}. When $E = C_L$ the Lagrangian points exist precisely but at $E > C_L$ an unstable Lyapunov periodic orbit is located close to each of these two points \citep[e.g.,][]{H69}. The Lagrangian points are saddle points of the effective potential, so when $E > C_L$, a star must pass close enough to one of these points in order to escape. Thus, in the case of a tidally limited star cluster the escape criterion is purely geometric. In particular, escapers are defined to be those stars moving in orbits beyond the tidal radius thus passing one of the two Lagrangian points with velocity pointing outwards \citep{FH00}.

The configuration and the phase space are divided into the bounded and unbounded regions. Usually, the vast majority of the bounded space is occupied by initial conditions of regular orbits forming stability islands. In many systems however, trapped chaotic orbits have also been observed. Therefore, we decided to distinguish between regular and chaotic trapped orbits. Over the years, several chaos indicators have been developed in order to determine the character of orbits. In our case, we choose to use the Smaller ALingment Index (SALI) method. The SALI \citep{S01} has been proved a very fast, reliable and effective tool, which is defined as
\begin{equation}
\rm SALI(t) \equiv min(d_-, d_+),
\label{sali}
\end{equation}
where $d_- \equiv \| {\bf{w_1}}(t) - {\bf{w_2}}(t) \|$ and $d_+ \equiv \| {\bf{w_1}}(t) + {\bf{w_2}}(t) \|$ are the alignments indices, while ${\bf{w_1}}(t)$ and ${\bf{w_2}}(t)$, are two deviation vectors which initially point in two random directions. For distinguishing between ordered and chaotic motion, all we have to do is to compute the SALI along time interval $t_{max}$ of numerical integration. In particular, we track simultaneously the time-evolution of the main orbit itself as well as the two deviation vectors ${\bf{w_1}}(t)$ and ${\bf{w_2}}(t)$ in order to compute the SALI. The time-evolution of SALI strongly depends on the nature of the computed orbit since when an orbit is regular the SALI exhibits small fluctuations around non zero values, while on the other hand, in the case of chaotic orbits the SALI after a small transient period it tends exponentially to zero approaching the limit of the accuracy of the computer $(10^{-16})$. Therefore, the particular time-evolution of the SALI allow us to distinguish fast and safely between regular and chaotic motion. Nevertheless, we have to define a specific numerical threshold value for determining the transition from order to chaos. After conducting extensive numerical experiments, integrating many sets of orbits, we conclude that a safe threshold value for the SALI is the value $10^{-8}$. In order to decide whether an orbit is regular or chaotic, one may follow the usual method according to which we check after a certain and predefined time interval of numerical integration, if the value of SALI has become less than the established threshold value. Therefore, if SALI $\leq 10^{-8}$ the orbit is chaotic, while if SALI $ > 10^{-8}$ the orbit is regular thus making the distinction between regular and chaotic motion clear and beyond any doubt. For the computation of SALI we used the \verb!LP-VI! code \citep{CMD14}, a fully operational routine which efficiently computes a suite of many chaos indicators for dynamical systems in any number of dimensions.

In our calculations, we set $10^4$ time units as a maximum time of the numerical integration. The vast majority of orbits (regular and chaotic) however, need considerable less time to find one of the two exits in the limiting surface and eventually escape from the system (obviously, the numerical integration is effectively ended when an orbit passes through one of the escape channels and escapes). Nevertheless, we decided to use such a vast integration time just to be sure that all orbits have enough time in order to escape. Remember, that there are the so called ``sticky orbits" which behave as regular ones during long periods of time (see also Chapter 47 in \citet{C04}). The SALI method can easily identify sticky orbits. In fact sticky orbits are those which hold intermediate values of SALI $(10^{-4} < $ SALI $ < 10^{-8})$ for long time intervals (see e.g., \citet{SABV04}). It should be clarified, that orbits which do not escape after a numerical integration of $10^4$ time units are considered as non-escaping or trapped regarding the value of SALI. In fact, orbits with escape periods larger than $10^4$ time units are completely irrelevant to our investigation since they lack physical meaning.

\section{Numerical results}
\label{numres}

In our investigation we seek to determine which orbits escape and which remain trapped, distinguishing simultaneously between regular and chaotic bounded motion\footnote{Generally, any dynamical method requires a sufficient time interval of numerical integration in order to distinguish safely between ordered and chaotic motion. Therefore, if the escape period of an orbit is very low or even worse if the orbit escapes directly from the system then, any chaos indicator will fail to work properly due to insufficient integration time. Hence, it is pointless to speak of regular or chaotic escaping orbits.}. Moreover, two additional aspects of the orbits will be examined: (i) the channel or exit through which a star escapes and (ii) the time-scale of the escape (we shall also use the terms escape period or escape rate). In particular, we shall explore these dynamical quantities for various values of the total orbital energy, always within the interval $\widehat{C} \in [0.001,0.1]$. As in Paper I, we shall deal only with the two-dimensional (2D) case of the star cluster model, therefore everywhere $z = \dot{z} = 0$.

\begin{figure*}
\centering
\resizebox{0.80\hsize}{!}{\includegraphics{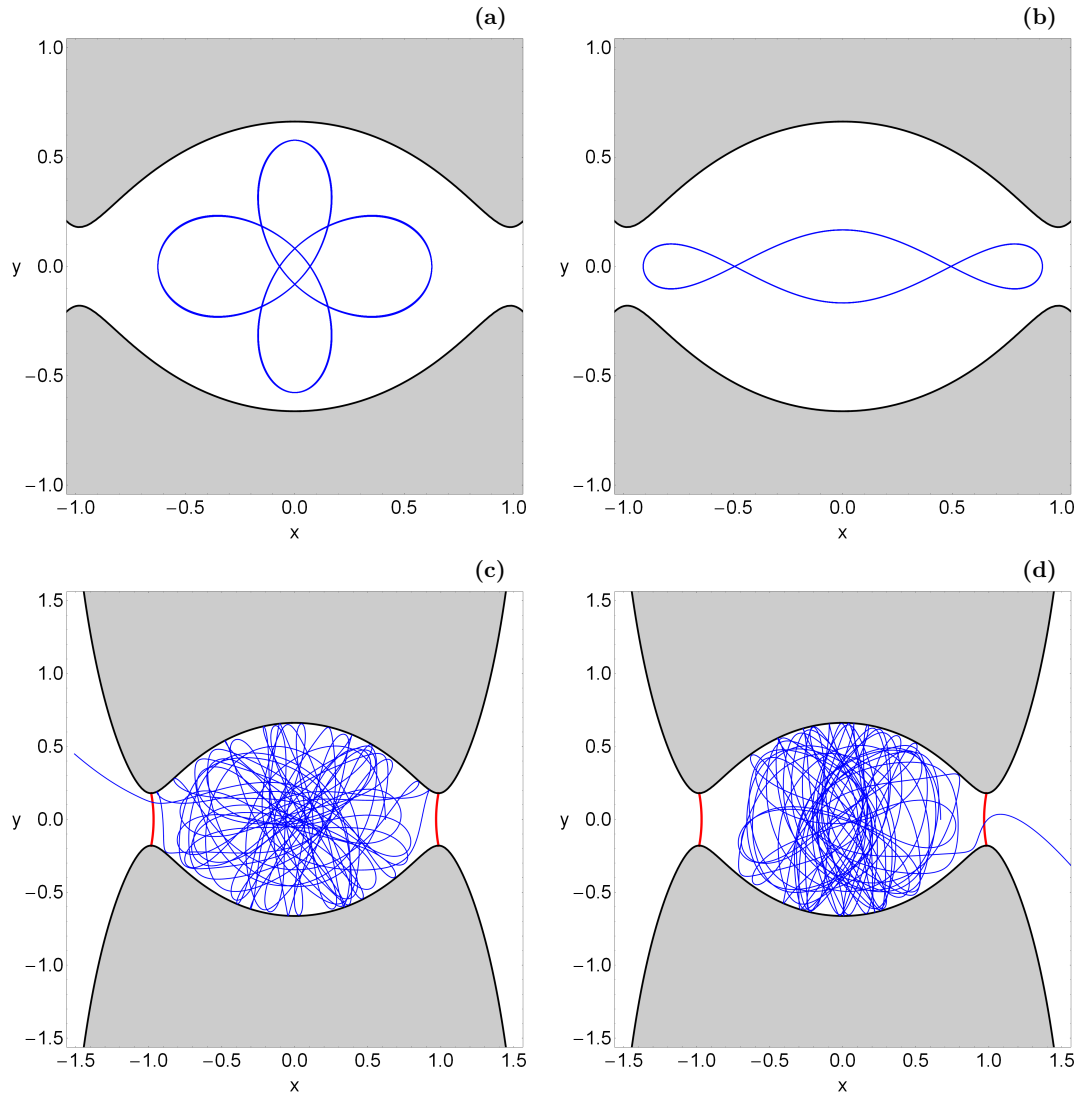}}
\caption{Characteristic examples of orbits in the star cluster where the isochrone potential is used. (a): 3:3 resonant orbit, (b): 1:3 resonant orbit, (c): orbit escaping through $L_1$, (d): orbit escaping through $L_2$. The red lines near the Lagrangian points denote the unstable Lyapunov orbits.}
\label{orbs}
\end{figure*}

Our preliminary analysis suggests that apart from the escaping orbits there is also a considerable amount of bounded orbits. In general terms, the majority of non-escaping bounded regions corresponds to initial conditions of regular orbits, where the adelphic integral of motion is present, restricting their accessible phase space and therefore hinders their escape. However, there are also chaotic orbits which do not escape within the predefined interval of $10^4$ time units and remain trapped for vast periods until they eventually escape to infinity. Therefore, we decided to classify the initial conditions of orbits in both the configuration and the phase space into four main categories: (i) orbits that escape through $L_1$, (ii) orbits that escape through $L_2$, (iii) non-escaping bounded regular orbits and (iv) trapped chaotic orbits. Additional numerical computations reveal that the non-escaping regular orbits are mainly loop orbits\footnote{Orbits which form a rosette or a simple circular closed path are known as loop orbits (see also Chapter 2 in \citet{C02}).} for which the adelphic integral applies, while other types of secondary resonant orbits are also present. In Fig. \ref{orbs}a we present an example of a 3:3 resonant orbit, while in Fig. \ref{orbs}b a characteristic example of a secondary 1:3 resonant orbit is shown. The $n:m$ notation we use for the regular orbits is according to \citet{CA98} and \citet{ZC13}, where the ratio of those integers corresponds to the ratio of the main frequencies of the orbit, where main frequency is the frequency of greatest amplitude in each coordinate. Main amplitudes, when having a rational ratio, define the resonances of an orbit. Finally in Figs. \ref{orbs}c-d we observe two orbits escaping through $L_1$ (exit channel 1) and $L_2$ (exit channel 2), respectively. The unstable Lyapunov orbits near the Lagrangian points are shown in red. Both regular orbits shown in Figs. \ref{orbs}a-b are computed until $t = 100$ time units, while on the other hand the escaping orbits presented in Figs. \ref{orbs}c-d were calculated for 5 time units more than the corresponding escape period in order to visualize better the escape trail. All orbits have $y_0 = \dot{x_0} = 0$, while the value of $\dot{y_0}$ is obtained from the Jacobian integral where $\widehat{C} = 0.01$. In Table \ref{table1} we provide the type, the exact initial condition $x_0$ and the escape period for all the depicted orbits. For modelling the star cluster the isochrone potential was used, while for the case of the Hernquist potential the types of orbits are similar. In Fig. \ref{orbs} we observe the two openings (exit channels) through which the particles can leak out. In fact, we may say that these two exits act as hoses connecting the interior region of the star cluster where $-r_t < x < r_t$ with the ``outside world" of the exterior region. Exit channel 1 (negative $x$-direction) indicates escape towards the galactic center, while channel 2 (positive $x$-direction) indicates escape towards infinity. The forbidden regions of motion within the banana-shaped isolines are shown in the same figure with gray.

\begin{figure*}
\centering
\resizebox{\hsize}{!}{\includegraphics{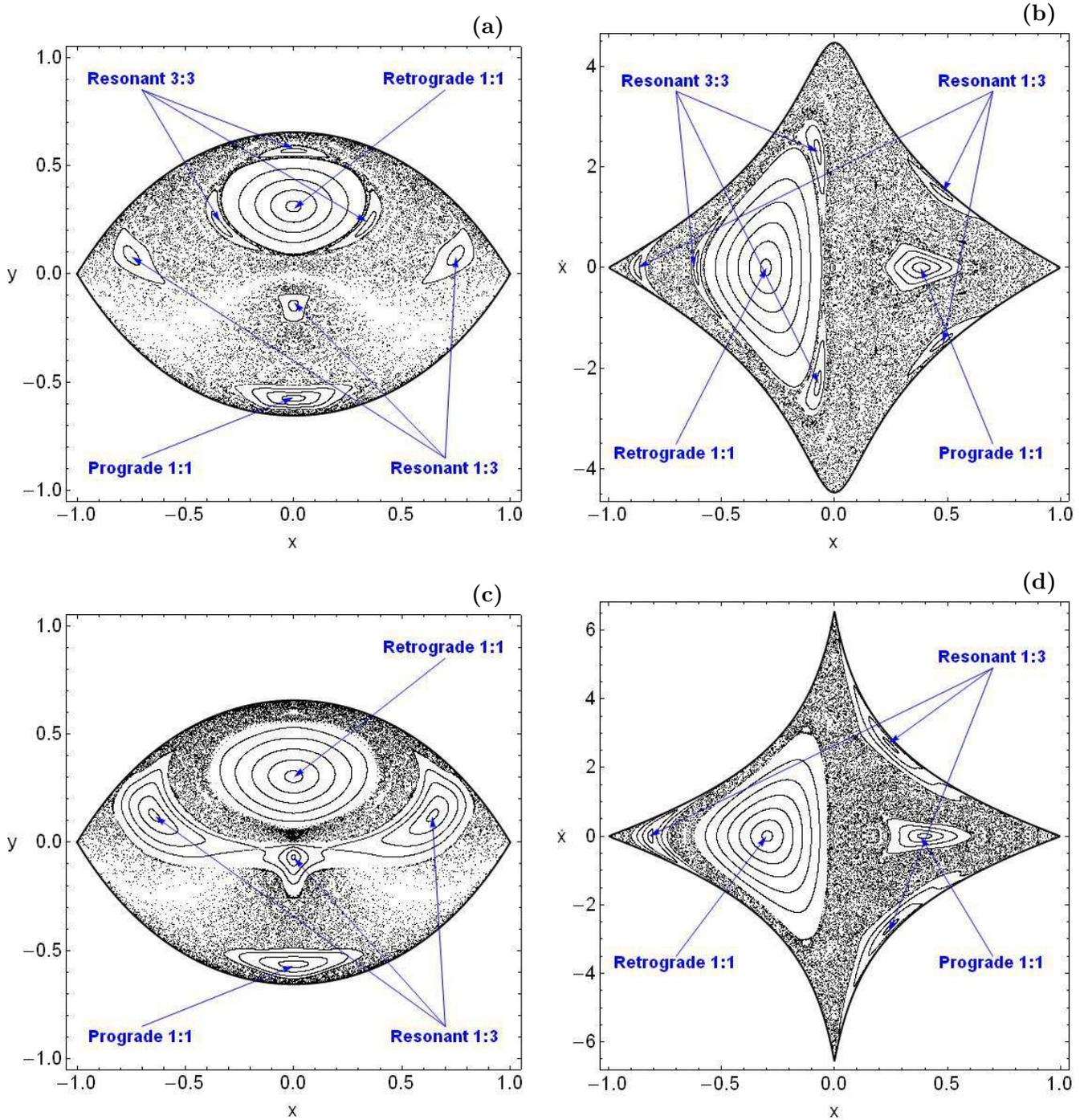}}
\caption{Poincar\'{e} Surfaces of Section (PSS) at the critical Jacobi energy $C_L$ for (first row): the isochrone potential and (second row): the Hernquist potential. (a and c left): The configuration $(x,y)$ space where orbits crossing $\dot{y} = 0$ with $\dot{x} > 0$ and (b and d right): the phase $(x,\dot{x})$ space where orbits crossing $y = 0$ with $\dot{y} > 0$. All types of regular orbits are identified.}
\label{pss}
\end{figure*}

\begin{table}
\centering
\setlength{\tabcolsep}{20pt}
\begin{center}
   \centering
   \caption{Type, initial conditions, and integration time $(t_{\rm int})$ of the orbits shown in Fig. \ref{orbs}(a-d).}
   \label{table1}
   \begin{tabular}{@{}lccccccr}
      \hline
      Figure & Type & $x_0$ & $t_{\rm int}$  \\
      \hline
      \ref{orbs}a & 3:3      & -0.625 & 100 \\
      \ref{orbs}b & 1:3      & -0.911 & 100 \\
      \ref{orbs}c & chaotic  &  0.620 &  76 \\
      \ref{orbs}d & chaotic  &  0.670 &  79 \\
      \hline
   \end{tabular}
\end{center}
\end{table}

A simple qualitative way for distinguishing between regular and chaotic motion in Hamiltonian systems is by plotting the successive intersections of the orbits using a Poincar\'{e} Surface of Section (PSS) \citep{HH64}. In particular, a PSS is a two-dimensional (2D) slice of the entire four-dimensional (4D) phase space. The following Fig. \ref{pss}(a-d) shows the PSSs for both potentials at the critical Jacobi energy $C_L$. Fig. \ref{pss}a corresponds to the configuration $(x,y)$ space showing orbits crossing $\dot{y} = 0$ with $\dot{x} > 0$, while orbits crossing $y = 0$ with $\dot{y} > 0$ at the phase $(x,\dot{x})$ space are presented in Fig. \ref{pss}b for the isochrone potential. Things are similar in Figs. \ref{pss}(c-d) for the Hernquist potential. In both cases, the initial conditions of the orbits are integrated forwards in time plotting a dot at each crossing through the surface of section. The outermost black thick curve circumscribing each PSS is the limiting curve which in the configuration space is defined as $\Phi_{\rm eff}(x,y,z=0) = E$, while in the phase space is given by
\begin{equation}
f(x,\dot{x}) = \frac{1}{2}\dot{x}^2 + \Phi_{\rm eff}(x, y=0, z=0) = E.
\label{zvc2}
\end{equation}
It is seen, that the vast majority of the surfaces of section is covered by a unified and extended chaotic sea however, stability regions with initial conditions corresponding to regular orbits are also present. In these stability islands the quasi-periodic orbits admit a third integral of motion, also known as ``adelphic integral" (see e.g., \citet{C63}), which hinders the test particles (stars) from escaping. Moreover, we observe that some areas in the chaotic sea of the $(x,y)$ plane are less densely occupied than others. We remark that in the following grid exploration in the configuration and phase space, we do not consider all initial conditions in the same chaotic domain as one the same orbit.

\subsection{The configuration $(x,y)$ space}
\label{cas1}

\begin{figure*}
\centering
\resizebox{0.8\hsize}{!}{\includegraphics{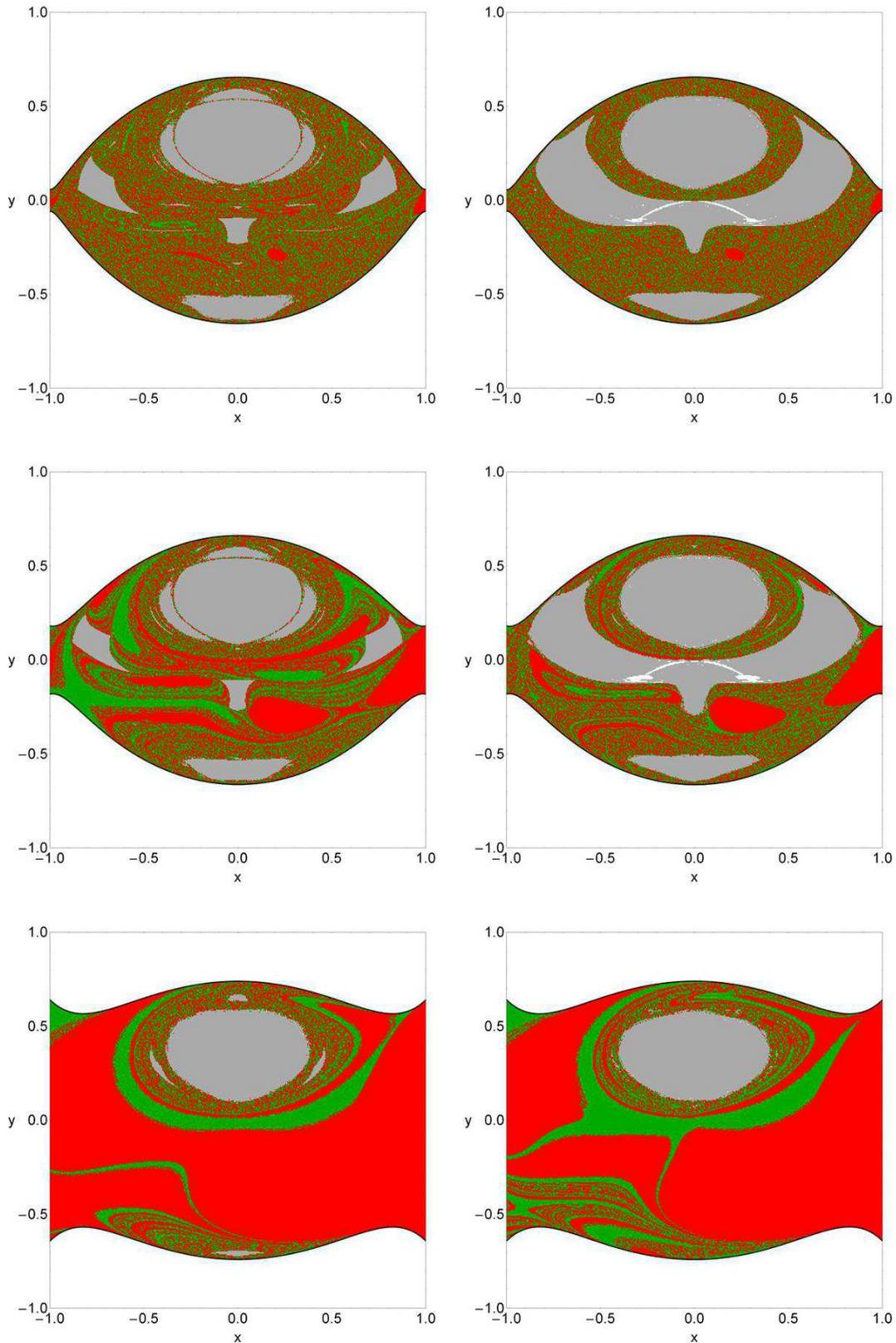}}
\caption{Orbital structure of the configuration $(x,y)$ space. Left column: case of the isochrone potential; Right column: case of the Hernquist potential. Top row: $\widehat{C} = 0.001$; Middle row: $\widehat{C} = 0.01$; Bottom row: $\widehat{C} = 0.1$. The green regions correspond to initial conditions of orbits where the stars escape through $L_1$, red regions denote initial conditions of orbits where the stars escape through $L_2$, gray areas represent stability islands of regular bounded orbits, while initial conditions of trapped chaotic orbits are shown in white.}
\label{grd1}
\end{figure*}

Our exploration begins in the configuration $(x,y)$ space. Fig. \ref{grd1} shows the orbital structure of the configuration space for three energy levels, where the four basic types of orbits are indicated with different colors. Specifically, gray color corresponds to regular non-escaping orbits, white color corresponds to trapped chaotic orbits, green color corresponds to orbits escaping through channel 1, while the initial conditions of orbits escaping through exit channel 2 are marked with red color. In the left column we present the results where the isochrone potential is used, while in the right column we see the outcomes of the Hernquist potential. Our results will be compared with the corresponding ones of the Plummer model shown in the right column of Fig. 5 of Paper I\footnote{In Paper I the total mass of the star cluster was $M_{\rm cl} = 2.2$, while in the cases of the isochrone and the Hernquist potential the corresponding value is $M_{\rm cl} = 2.7$. However, our calculations suggest that this minor deviation in the total mass of the star cluster does not significantly affect the orbital structure. Therefore, we decided not to reconstruct the orbital diagrams of the Plummer case but instead use those of Paper I.}. We observe that for all energy levels there are many similarities between the three models but there are also significant differences. In particular, it is seen that the main stability islands containing the initial conditions of all the 1:1 resonant orbits are almost the same everywhere. However the structure of the secondary stability islands is very different. Specifically one can see that in both the isochrone and the Hernquist potentials the set of four small islands of the Plummer case corresponding to the 4:4 resonant orbits which are located around the lower 1:1 region is now absent. Moreover, the set of the 3:3 stability islands around the upper 1:1 region is absent in the case of the Hernquist potential. Undoubtedly, the most striking difference concerns the 1:3 stability regions. In particular, in the isochrone case their domains are more extended than in the case of the Plummer potential. The 1:3 stability regions are much more extended in the Hernquist case where we observe that the three islands are almost connected to each other. With a more closer look we can distinguish a thin layer of initial conditions of trapped chaotic orbits separating the 1:3 regions. When $\widehat{C} = 0.001$ in all thee cases the vast majority of the $(x,y)$ plane is highly fractal, while for higher energy levels the fractality of the plane is reduced and well-formed basins of escape dominate. When $\widehat{C} = 0.1$ (the highest energy level studied) the 1:3 stability islands have disappeared and the vast majority of the configuration plane is occupied by several basins of escape. In all three cases the structure of the escape basins is quite similar and only small differences can be mentioned (i.e, the fractal structures at the lower part $(y < 0)$ of the configuration plane are more prominent in the Hernquist case when $\widehat{C} = 0.1$).

\begin{figure*}
\centering
\resizebox{\hsize}{!}{\includegraphics{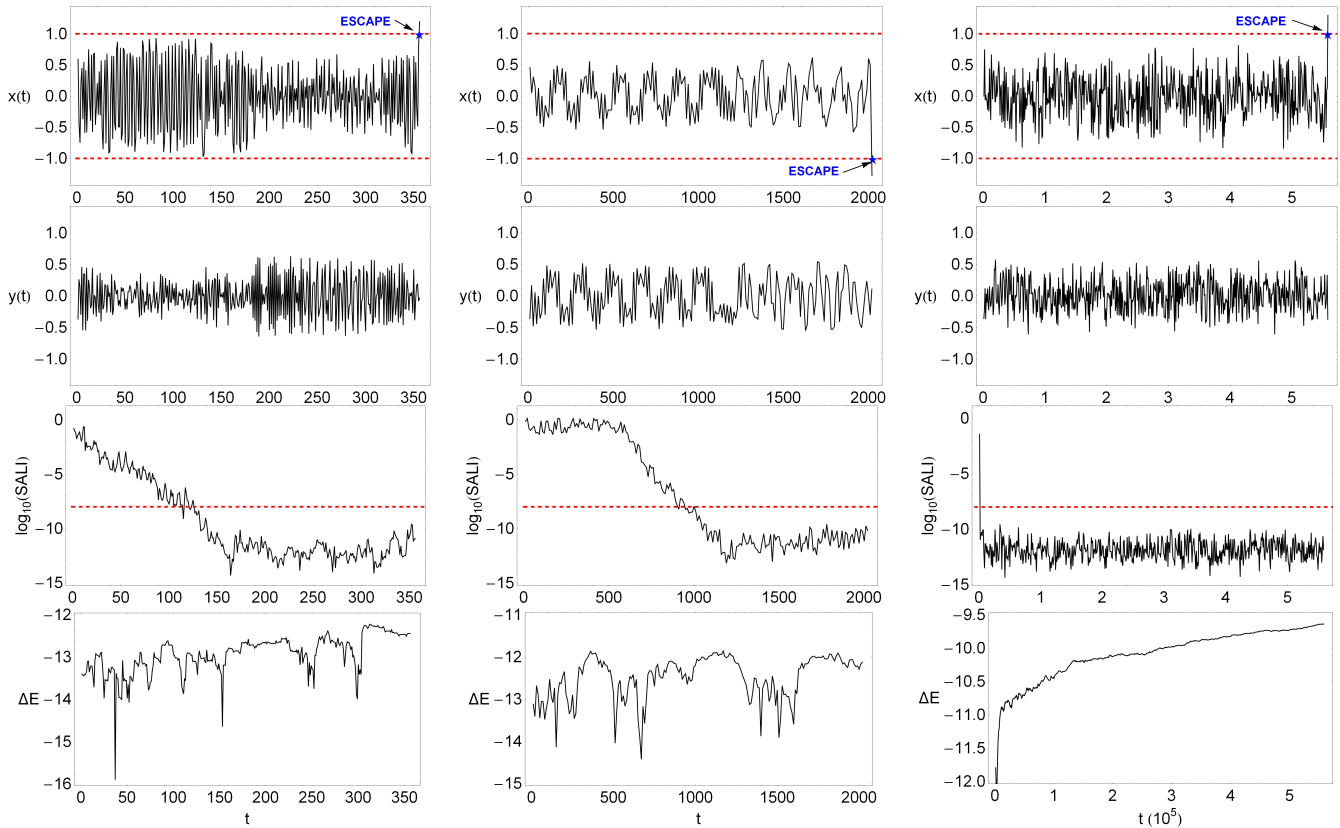}}
\caption{Time-evolution of $x$ and $y$ coordinates, the SALI and the integration error in the Jacobian integral of (first column): a fast escaping orbit; (second column): a sticky orbit; (third column): a trapped chaotic orbit in the case of the Hernquist potential.}
\label{tevol}
\end{figure*}

\begin{figure*}
\centering
\resizebox{0.8\hsize}{!}{\includegraphics{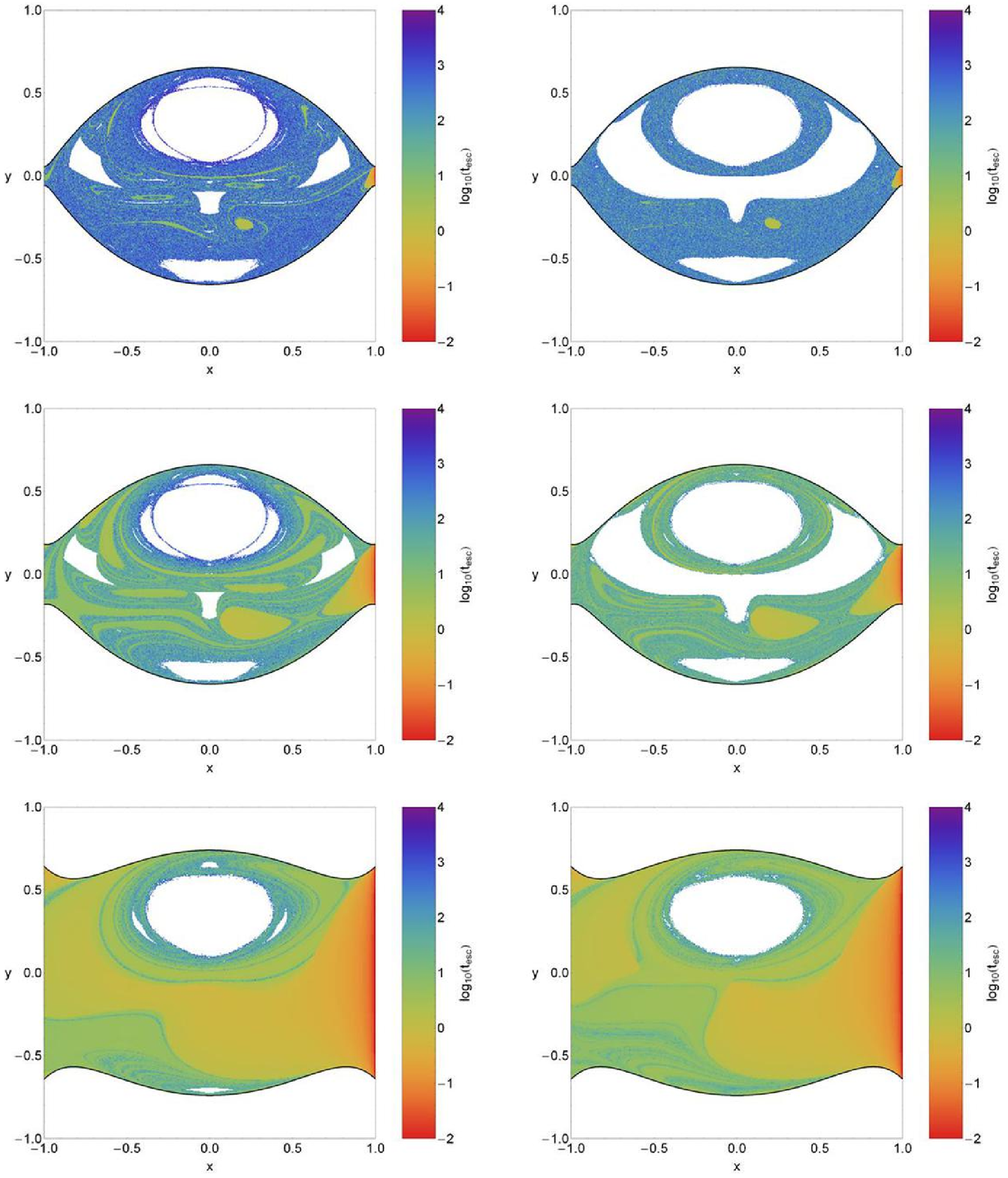}}
\caption{Distribution of the escape times $t_{\rm esc}$ of the orbits on the configuration $(x,y)$ space. Left column: case of the isochrone potential; Right column: case of the Hernquist potential. Top row: $\widehat{C} = 0.001$; Middle row: $\widehat{C} = 0.01$; Bottom row: $\widehat{C} = 0.1$. The darker the color, the larger the escape time. Initial conditions of non-escaping regular orbits and trapped chaotic orbits are shown in white.}
\label{tesc1}
\end{figure*}

\begin{figure*}
\centering
\resizebox{0.8\hsize}{!}{\includegraphics{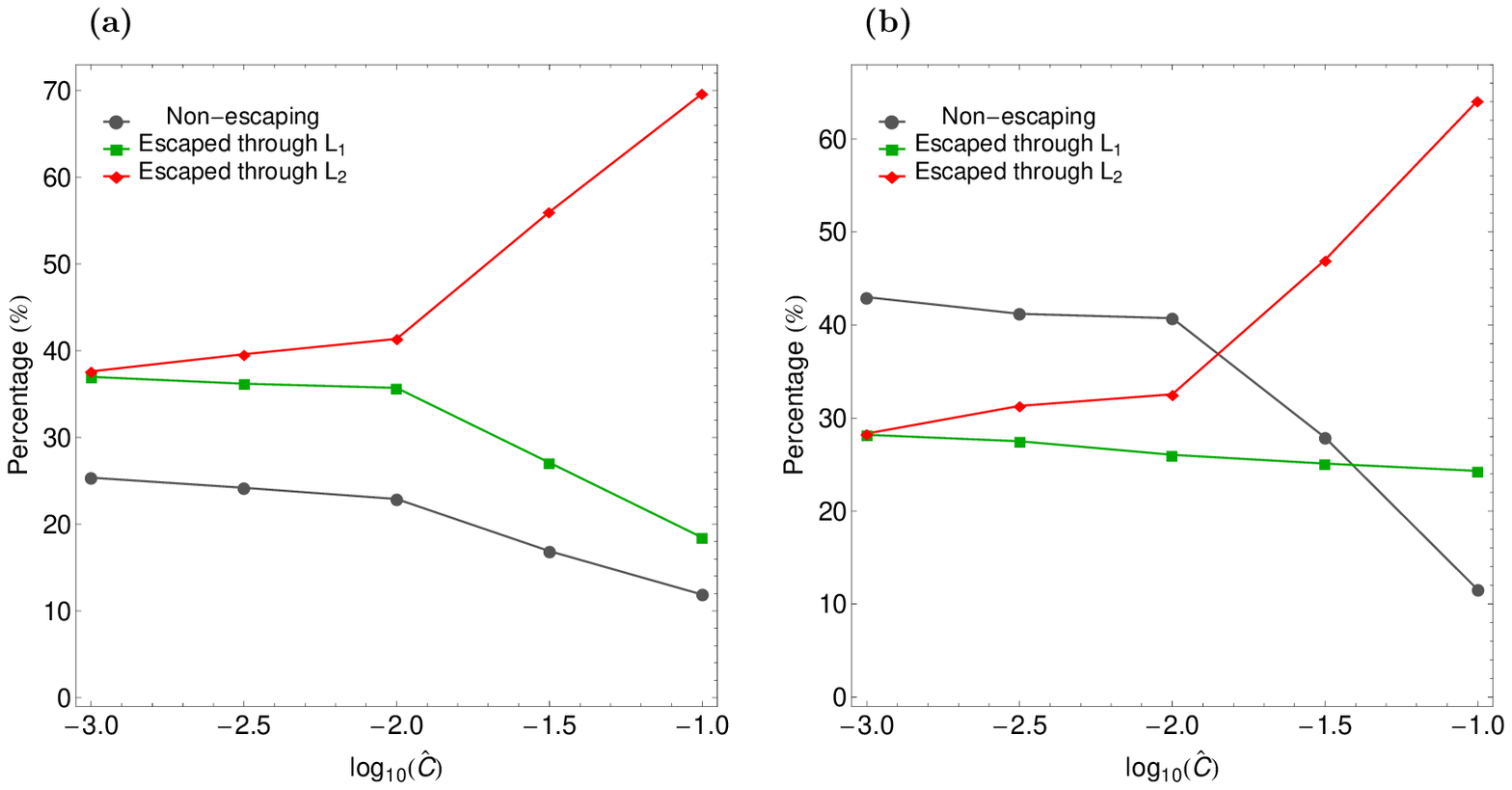}}
\caption{Evolution of the percentages of escaping and non-escaping orbits when varying the energy parameter $\widehat{C}$ on the configuration $(x,y)$ space for the case of the (a-left): isochrone potential and (b-right): Hernquist potential.}
\label{percs1}
\end{figure*}

The first column of Fig. \ref{tevol} shows the the time-evolution of the $x$ and $y$ coordinates of a fast escaping orbit in the Hernquist potential with initial conditions: $x_0 = -0.52$, $y_0 = 0$, $\dot{y_0} = 0$, while the value of $\dot{x_0}$ is obtained from the Jacobian integral (\ref{ham}) when $\widehat{C} = 0.001$. The horizontal dashed red lines in the upper panel denote the position of the two Lagrangian points delimiting the transition from trapped to escaping motion. In the same figure we see the time-evolution of SALI as well as the time-evolution of the error in the Jacobian integral $(\Delta E = (|E - E_0|)/E_0)$. We see that the orbit need only about 110 time units in order to cross the SALI threshold value, while it escapes from channel 2 after about 246 time units. The error in the Jacobian integral proves the accuracy of our numerical integrator. In the second column of Fig. \ref{tevol} we present the time-evolution of a stick orbit with initial conditions: $x_0 = 0$, $y_0 = 0.062$, $\dot{y_0} = 0$, while the value of $\dot{x_0}$ is obtained from the Jacobian integral (\ref{ham}) when $\widehat{C} = 0.001$. According to the time-evolution of SALI this orbit behave as regular (SALI $ > 10^{-4}$) for the first about 600 time units. Then in the interval $600 < t < 900$ it displays intermediate values of SALI (sticky period) before it escapes form channel 1 after about 2020 time units.

In the case of the Hernquist potential (see right column of Fig. \ref{grd1}) we identified some trapped chaotic orbits. At this point we would like to try to interpret and also justify the phenomenon of trapped chaos. By the term ``trapped chaos" we refer to the case where a chaotic orbit remains trapped for a vast time interval inside an open equipotential surface. It should be emphasized and clarified that these trapped chaotic orbits cannot be considered by no means neither as sticky orbits nor as super sticky orbits\footnote{A super sticky orbit is an orbit with a sticky period larger than $10^4$ time units. The sticky period is defined as the time interval in which SALI has intermediate values $(10^{-4} < $ SALI $ < 10^{-8})$ before it crosses the threshold value of $10^{-8}$.} with sticky periods larger than $10^4$ time units. Sticky orbits are those who behave regularly for long time periods before their true chaotic nature is fully revealed. In our case on the other hand, this type of orbits exhibit chaoticity very quickly as it takes no more than about 100 time units for the SALI to cross the threshold value (SALI $\ll 10^{-8}$), thus identifying beyond any doubt their chaotic character. Chaotic orbits do not admit a third integral of motion, therefore according to the classical chaos theory for time $t \rightarrow \infty$ they will fill all the available space inside the equipotential surface, unless they escape. In other words, given enough time of numerical integration all trapped chaotic orbits will eventually pass through an exit and escape from the star cluster. To prove this, we chose such an initial condition and let the time running. In the third column of Fig. \ref{tevol} we present the time-evolution of the $x$ and $y$ coordinates of an orbit with initial conditions: $x_0 = -0.37582418$, $y_0 = -0.10329670$, $\dot{y_0} = 0$, while the value of $\dot{x_0}$ is obtained from the Jacobian integral (\ref{ham}) when $\widehat{C} = 0.01$. This particular orbit is trapped chaotic and as we can see a huge time of numerical integration, about 558000 time units is required so that the orbit can escape from channel 2. It should be emphasized that for this orbit the SALI crosses the threshold value after about 150 time units and its values remain in the chaotic zone (SALI < $10^{-8}$) during the entire time interval before the escape. We observe in the last panel that the error in the Jacobian integral grows during the vast numerical integration but always in reasonable values. We experimented with several trapped chaotic orbits and the obtained results were very similar. Stars moving in chaotic orbits restricted inside the interior region of the cluster having large escape times have also been reported in previous investigations \citep[see e.g., Fig. 11 in][]{FH00}.

Perhaps the presence of a small percentage of trapped chaotic orbits in the case of the Hernquist potential is related to the presence of the cusp near the center of the potential where the density of Eq. (\ref{3den}) gets singular and the velocities become arbitrarily high. Additional numerical calculations indicate that the phenomenon of trapped chaos is not an artifact of the numerical integration since the areas covered by trapped chaotic orbits remain unperturbed when we use smaller time step (0.005 or 0.001) in the numerical integration. One may suspect that the trapped chaotic orbits might pass close to the origin at $(x,y) = (0,0)$, where the singular cusp is located and they might escape relatively fast because of an energy error in the Jacobian since the velocities can become relatively large near the singularity. Our numerical calculations however, indicate that this scenario is not plausible. These orbits do eventually escape but only after vast time intervals of more than $10^{5}$ time units.

Another possible reason for the phenomenon of trapped chaos observed in the Hernquist model might be the force acting on stars near the center of the star cluster. For both the Plummer and the isochrone potentials the force near the center vanishes, while in the case of the Hernquist potential on the other hand it is non zero. In particular, according \citet{H90}
\begin{equation}
F(r) = - \frac{d\Phi}{dr} = - \frac{G M_{\rm cl}}{\left(r + r_{\rm H}\right)^2}.
\end{equation}
For $r = 0$ we have that $F(r = 0) = -232.242$.

In the following Fig. \ref{tesc1} we present how the escape times $t_{\rm esc}$ of orbits are distributed on the configuration $(x,y)$ space for both potentials. Light reddish colors correspond to fast escaping orbits with short escape periods, dark blue/purple colors indicate large escape rates, while white color denote both non-escaping regular and trapped chaotic orbits. It is evident, that orbits with initial conditions close to the boundaries between the escape basins, that is, the fractal areas of the plots need significant amount of time in order to escape from the cluster, while on the other hand, inside the basins of escape where there is no dependence on the initial conditions whatsoever, we measured the shortest escape rates of the orbits. We observe that for $\widehat{C} = 0.001$ the escape periods of orbits with initial conditions in the fractal basin boundaries are huge corresponding to tens of thousands of time units. This phenomenon is anticipated because in this case the width of the escape channels is very small and therefore, the orbits should spend much time inside the equipotential surface until they find one of the two openings and eventually escape to infinity. One may observe, that as the value of the energy increases however, the escape channels become more and more wide, while the escape times of all orbits (inside the exit basins and in the fractal regions) is reduced considerably. Our calculations suggest that statistically the escape times in the case of the Hernquist potential are lower (almost half) than in the case where the star cluster is modelled by the isochrone potential. Moreover, we see that for $\widehat{C} = 0.01$ and $\widehat{C} = 0.1$ the escaping orbits with initial conditions near the boundaries of the stability islands have high escape times (more than 1000 time units). This is true because near the boundaries of the stability islands is usually the location of sticky orbits which need a sufficient time interval before they reveal their true chaotic nature and finally escape from the system (see second column of Fig. \ref{tevol}).

The evolution of the percentages of all types of orbits on the configuration space for both potentials when the dimensionless energy parameter $\widehat{C}$ varies is presented in Fig. \ref{percs1}(a-b). Due to an unreasonably difficulty in the data analysis we decided to investigate in the configuration space (similar approach in the following phase space) only five different values of $\widehat{C}$. The case of the isochrone potential is shown in Fig. \ref{percs1}a. When $\widehat{C} = 0.001$, that is an energy level just above the escape energy, about one fourth of the total plane is covered by initial conditions of non-escaping regular orbits, while the rest 75\% is devoted to escaping orbits. In particular the rates of both channels are identical suggesting that in this case the configuration plane is highly fractal. As we proceed to higher energy levels the percentage of the non-escaping regular orbits is reduced and when $\widehat{C} = 0.1$ is only about 12\%. On the contrary we observe that always the escaping orbits is the most populated family. With increasing energy the rates of both channels (exits) start to diverge and especially for $\widehat{C} > 0.01$ the percentage of escapers through $L_1$ decreases, while that of escaping orbits through $L_2$ increases linearly. At the highest energy level studied $(\widehat{C} = 0.1)$, escaping orbits through channel 1 correspond to about 18\% of the $(x,y)$ plane, while escapers through channel 2 dominate covering the rest 70\%. In the case of the Hernquist potential (Fig. \ref{percs1}b) the profile of the curves is similar however there are some important differences: (i) in the interval $0.001 < \widehat{C} < 0.01$ non-escaping regular orbits is the most populated type of orbits; (ii) the percentage of escaping orbits trough channel 1 exhibits only a minor decrease from 28\% to 24\%; (iii) escaping orbit through $L_2$ dominate only for $\widehat{C} > 0.01$. At this point we should point out that in both cases trapped chaotic orbits possess always a very weak percentage (less than 1\%) and this is the main reason why this evolution of this type is not included in the diagrams. Thus taking into account all the above-mentioned analysis we may reasonably conclude that in the Hernquist models there is a greater percentage of bounded regular orbits mainly due to the strong presence of the 1:3 family. Furthermore, in both cases for relatively high energy levels escaping orbits through exit 2 dominate.

\subsection{The phase $(x,\dot{x})$ space}
\label{cas2}

\begin{figure*}
\centering
\resizebox{0.8\hsize}{!}{\includegraphics{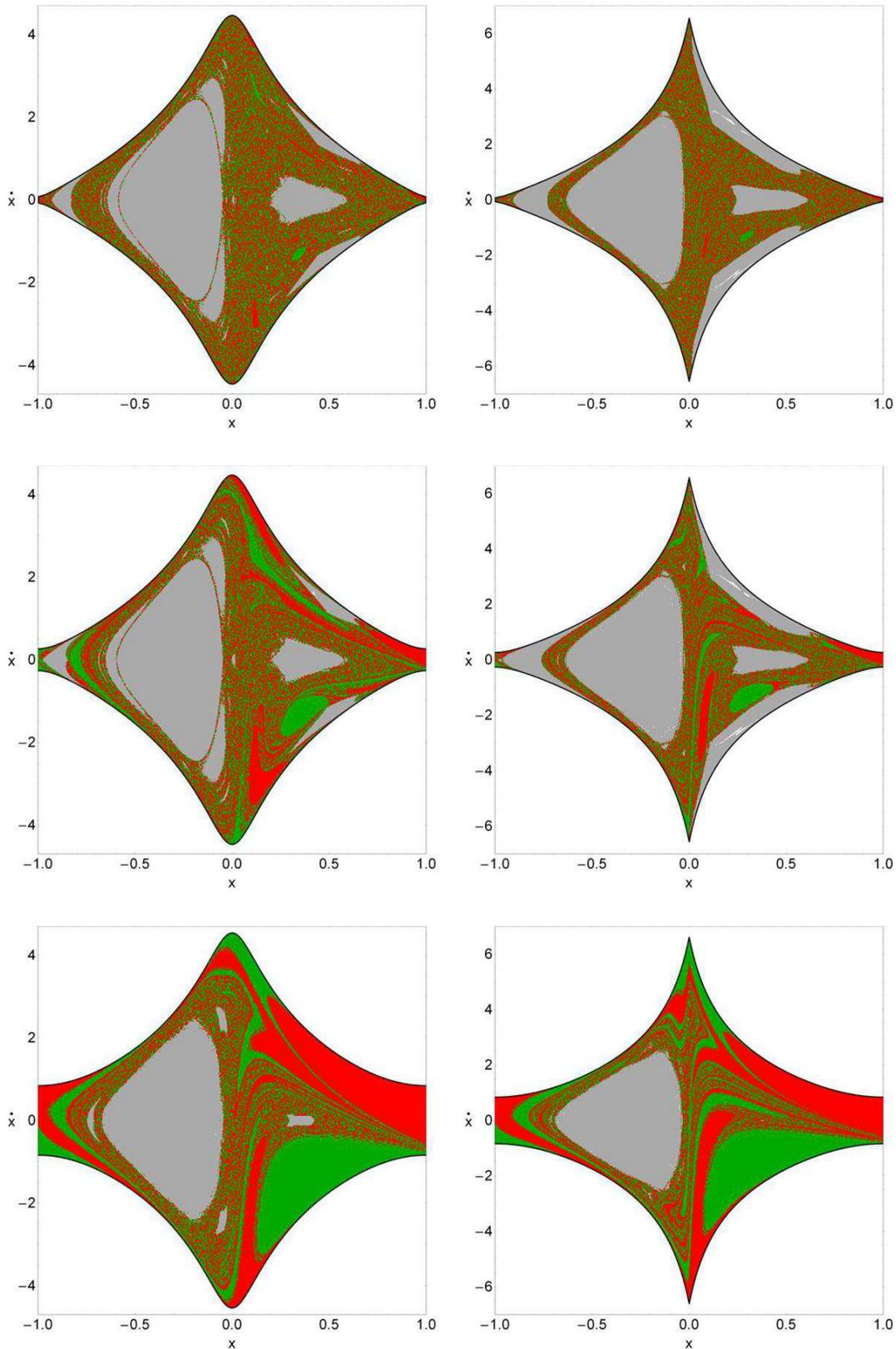}}
\caption{Orbital structure of the phase $(x,\dot{x})$ space. Left column: case of the isochrone potential; Right column: case of the Hernquist potential. Top row: $\widehat{C} = 0.001$; Middle row: $\widehat{C} = 0.01$; Bottom row: $\widehat{C} = 0.1$. The green regions correspond to initial conditions of orbits where the stars escape through $L_1$, red regions denote initial conditions of orbits where the stars escape through $L_2$, gray areas represent stability islands of regular bounded orbits, while initial conditions of trapped chaotic orbits are shown in white.}
\label{grd2}
\end{figure*}

\begin{figure*}
\centering
\resizebox{0.8\hsize}{!}{\includegraphics{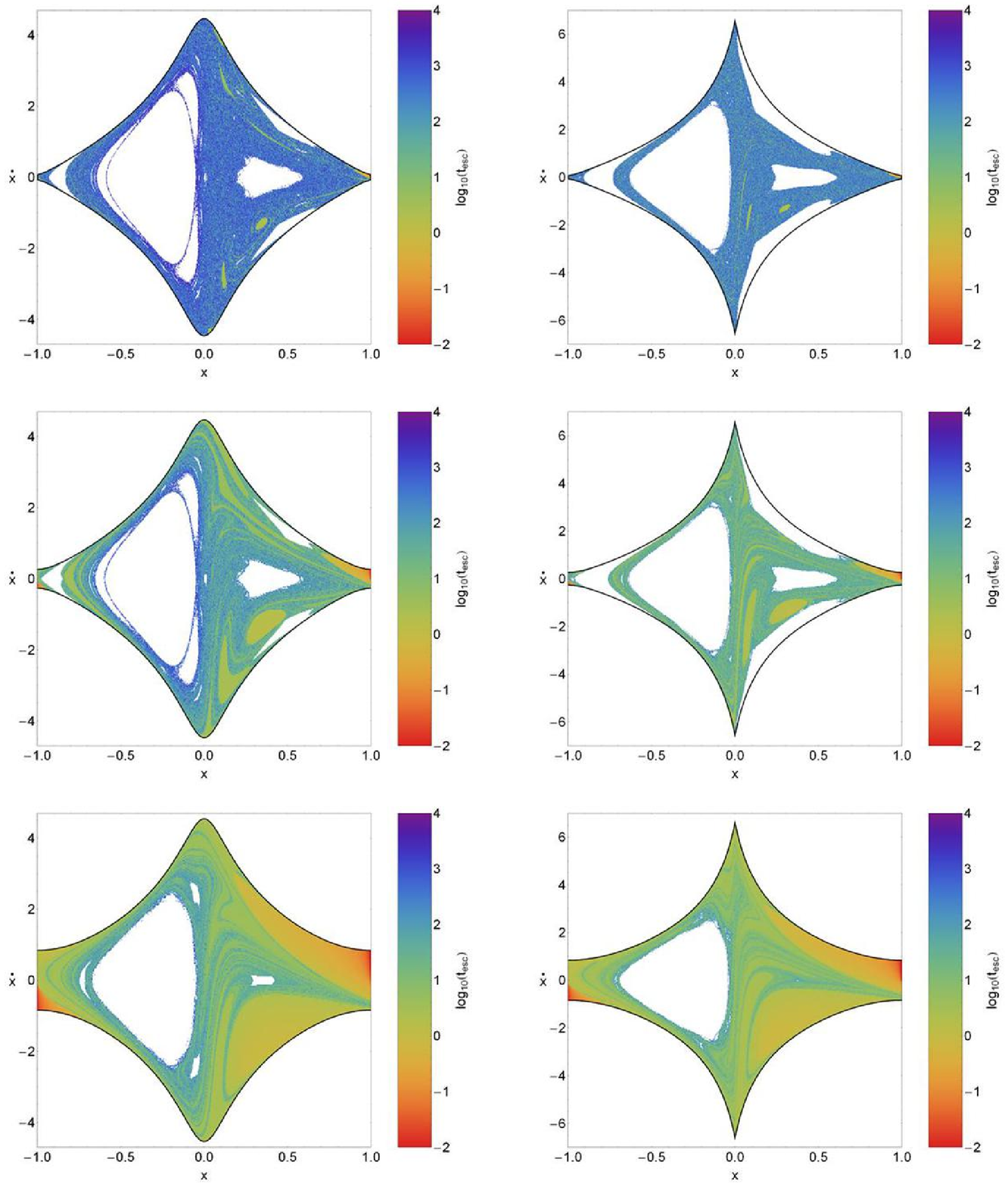}}
\caption{Distribution of the escape times $t_{\rm esc}$ of the orbits on the phase $(x,\dot{x})$ space. Left column: case of the isochrone potential; Right column: case of the Hernquist potential. Top row: $\widehat{C} = 0.001$; Middle row: $\widehat{C} = 0.01$; Bottom row: $\widehat{C} = 0.1$. The darker the color, the larger the escape time. Initial conditions of non-escaping regular orbits and trapped chaotic orbits are shown in white.}
\label{tesc2}
\end{figure*}

\begin{figure*}
\centering
\resizebox{0.8\hsize}{!}{\includegraphics{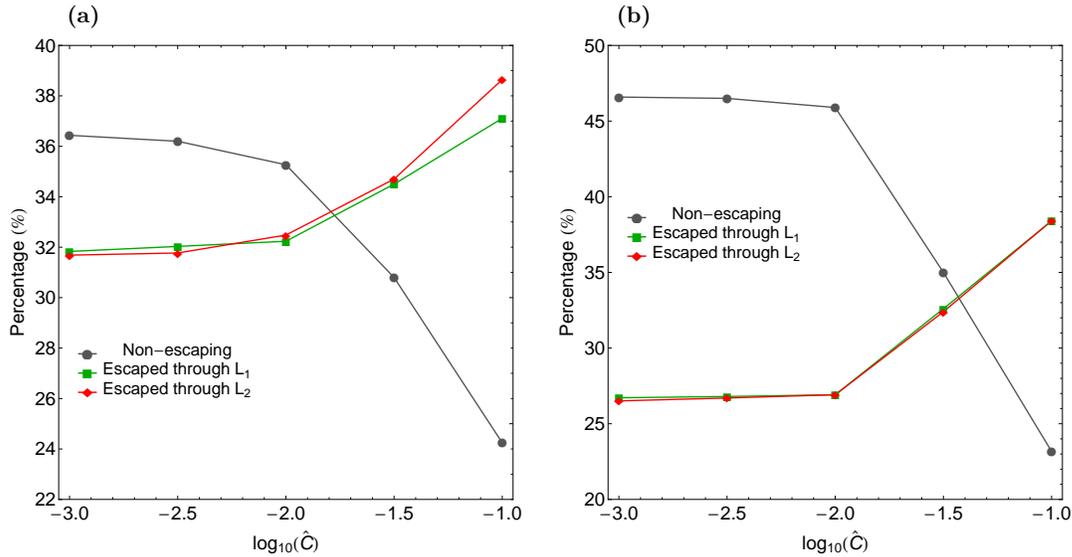}}
\caption{Evolution of the percentages of escaping and non-escaping orbits when varying the energy parameter $\widehat{C}$ on the phase $(x,\dot{x})$ space for the case of the (a-left): isochrone potential and (b-right): Hernquist potential.}
\label{percs2}
\end{figure*}

Our investigation continues in the phase $(x,\dot{x})$ space where we follow the same numerical approach as discussed previously. In Fig. \ref{grd2} we depict the orbital structure of the $(x,\dot{x})$ plane for the isochrone and Hernquist cases and for the three energy levels, using different colors in order to distinguish between the four main types of orbits (non-escaping regular; trapped chaotic; escaping through $L_1$ and escaping through $L_2$). This time our results will be compared with the corresponding ones of the Plummer model shown in the left column of Fig. 5 of Paper I. The observed regular areas correspond mainly to retrograde 1:1 resonant orbits \citep[see also][]{FH00,EGFJ07} (i.e., when a star revolves clockwise around the cluster in the opposite sense with respect to the motion of the cluster around the parent galaxy), while there is also a smaller stability island of prograde counterclockwise 1:1 resonant orbits in the right part $(x > 0)$ of the plane. As for the other resonant orbits the similarities and differences are the same with that explained in the previous subsection. However there is a new significant difference. In particular, we observe that in the case of the Hernquist potential stars can obtain much greater velocity near the center due to the power-law cusp of the potential. Regarding the structure of the fractal areas and the several escape basins we may say that in general terms there are no important differences between the three cases.

Here we must point out, that the phase $(x,\dot{x})$ plane shown in Fig. \ref{grd2} is not a classical PSS, simply because escaping orbits in general, do not intersect the $y = 0$ axis after a certain time, thus preventing us from defying a recurrent time. A classical PSS exists only if orbits intersect an axis, like $y = 0$, at least once within a certain time interval. Nevertheless, in the case of escaping orbits we can still define local surfaces of section which help us to understand the orbital behavior of the dynamical system.

We see in the $(x,\dot{x})$ surface of section shown in Fig. \ref{pss}d for the Hernquist model that the invariant curves around the 1:3 periodic orbit located on the $x$ axis block the exit close to $L_1$. Of course in this case the channel is closed due to the fact that the surface of section is at the critical value of the Jacobi integral. However when $E > C_L$ the stars moving in trapped chaotic orbit, after more than $10^5$ time units, escape between these invariant curves and the boundary of the surface of section on the top or bottom side and, after than, they reach the exist at $L_1$. Our computations suggest that trapped chaotic orbits exist only for $\widehat{C} = 0.001$ and $\widehat{C} = 0.01$ when the stability islands of the 1:3 resonance are present. On the other hand, for $\widehat{C} = 0.1$ where the 1:3 stability islands disappear we did not identify any initial conditions of trapped chaotic orbits. Therefore we may say that the presence of the 1:3 stability island may be responsible, in a way, for the observed trapped chaos. It should be noted however, that the 1:3 stability island is also present in the case of the isochrone model, where no trapped chaotic orbits were reported.

The distribution of the escape times $t_{\rm esc}$ of orbits on the phase space for both cases is shown in Fig. \ref{tesc2}. One may observe that the results are very similar to those presented earlier in Fig. \ref{tesc1}, where we found that orbits with initial conditions inside the escape basins have the smallest escape rates, while on the other hand, the longest escape times correspond to orbits with initial conditions in the fractal regions of the plots or in the boundaries of the stability islands. At this point, we would like to point out that the basins of escape can be easily distinguished being the regions with intermediate greenish colors indicating fast escaping orbits. Indeed, our numerical calculations suggest that orbits with initial conditions inside the basins need no more that 10 time units in order to escape from the star cluster.

Fig. \ref{percs2}(a-b) shows how the percentages of all types of orbits on the phase space evolve for both cases when the energy parameter varies in the interval $\widehat{C} \in [0.001,0.1]$. For the isochrone potential we observe in Fig. \ref{percs2}a that for low energy levels non-escaping regular orbits occupy about 36\% of the phase space. As the value of the energy increases however, the portion of the ordered orbits is reduced and when $\widehat{C} = 0.1$ it corresponds to only 24\% of the $(x,\dot{x})$ plane. It is also seen that the rates of the escaping orbits evolve almost identically and only at relatively high values of the energy the rates slightly diverge, being the escapers through exit 2 more populated. Things are quite similar in the case of the Hernquist potential. Indeed in Fig. \ref{percs2}b we see that in the interval $0.001 < \widehat{C} < 0.01$ bounded regular orbits cover about 46.5\% of the phase space, while for larger values of the energy there is a steep fall of the rate to about 23\% when $\widehat{C} = 0.1$. Furthermore, the rates of the escaping orbits are exactly the same everywhere. Initially escaping orbits share about 52\% of the phase space, while for $\widehat{C} > 0.01$ their rates exhibit a significant growth and at the highest energy level studied we found that 77\% of the phase space is covered by initial conditions of escaping orbits. Once again, as we found in the configuration space, the rate of trapped chaotic orbits is extremely low (always less than 1\%). Therefore one can argue that in the phase space both exit channels are equiprobable throughout the examined energy interval.

\subsection{An overview analysis}
\label{geno}

\begin{figure*}
\centering
\resizebox{\hsize}{!}{\includegraphics{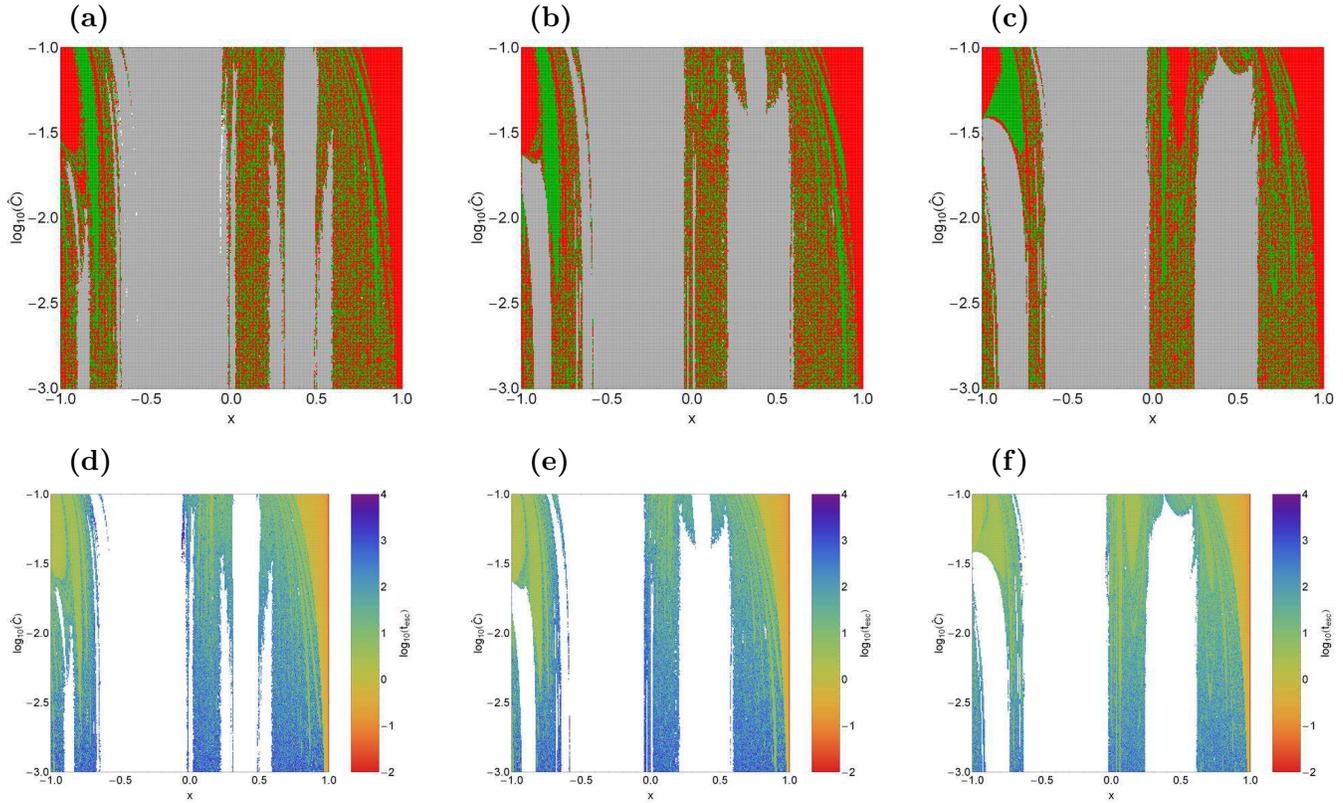}}
\caption{(first row): Orbital structure of the $(x,\widehat{C})$-plane; (second row): the distribution of the corresponding escape times of the 2D orbits. (first column): Plummer potential; (second column): Isochrone potential; (third column): Hernquist potential. The color codes are the exactly same as in Figs. \ref{grd1} and \ref{tesc1}.}
\label{xct}
\end{figure*}

\begin{figure*}
\centering
\resizebox{\hsize}{!}{\includegraphics{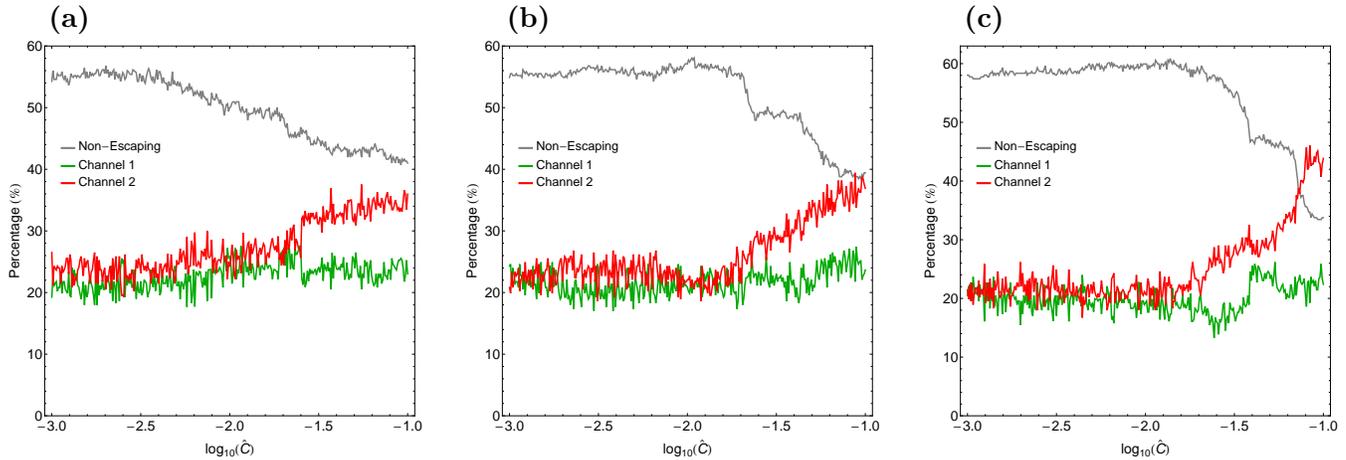}}
\caption{Evolution of the percentages of escaping and non-escaping orbits on the $(x,\widehat{C})$-plane as a function of the dimensionless energy parameter $\widehat{C}$ for the (a): Plummer potential; (b): isochrone potential; (c): Hernquist potential.}
\label{percs3}
\end{figure*}

\begin{figure*}
\centering
\resizebox{\hsize}{!}{\includegraphics{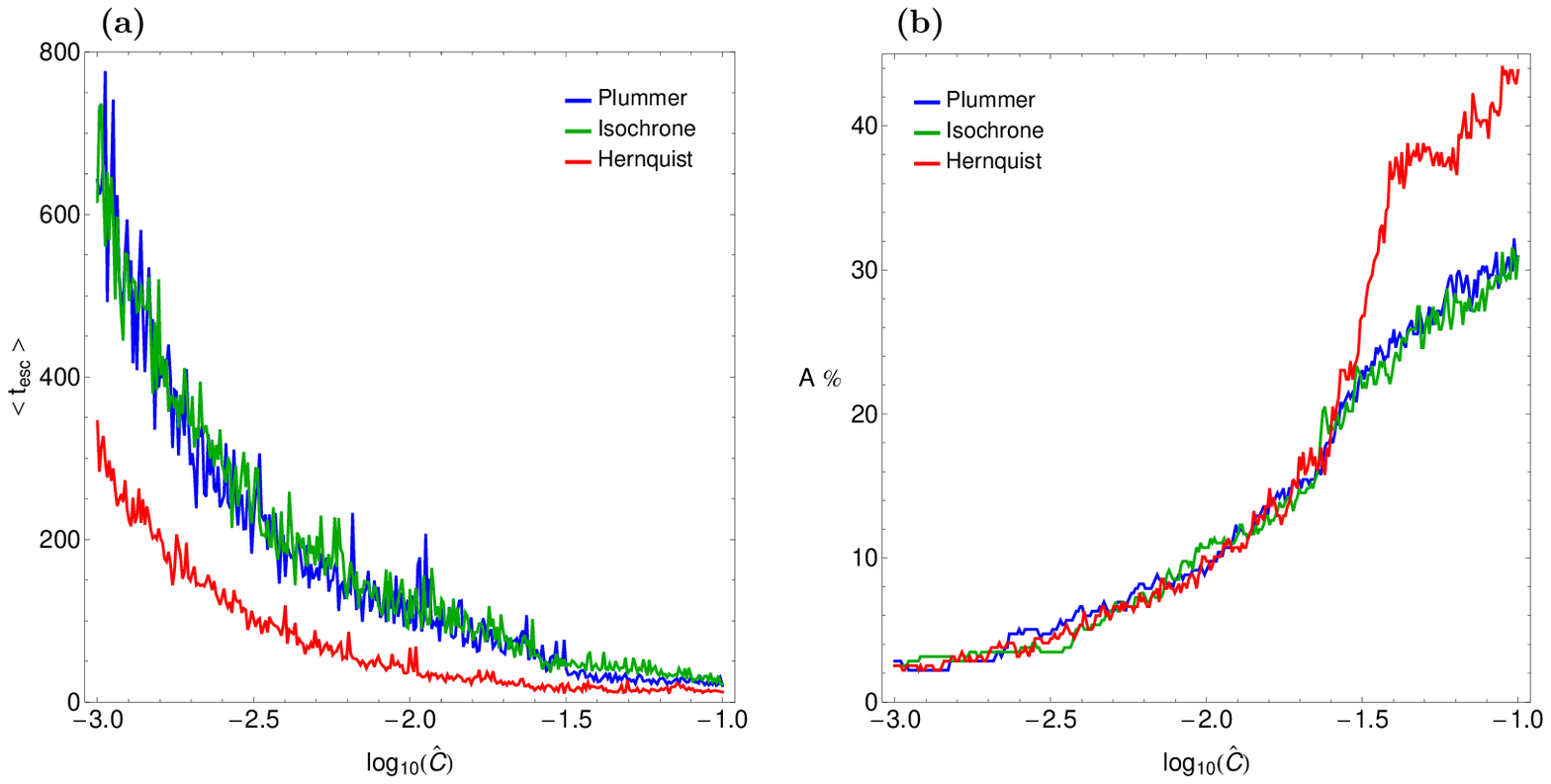}}
\caption{(a-left): The average escape time of orbits $< t_{\rm esc} >$ and (b-right): the percentage of the $(x,\widehat{C})$ planes covered by the escape basins as a function of the dimensionless energy parameter $\widehat{C}$ for all three cases.}
\label{stats}
\end{figure*}

The color-coded grids that discussed earlier in the configuration $(x,y)$ as well as in the phase $(x,\dot{x})$ space provide sufficient information on the phase space mixing for only a fixed value of the Jacobi constant. H\'{e}non however, back in the late 60s \citep[e.g.,][]{H69}, introduced a new type of plane which can provide information not only about stability and chaotic regions but also about areas of bounded and escaping orbits using the section $y = \dot{x} = 0$, $\dot{y} > 0$. In other words, all the 2D orbits of the stars of the cluster are launched from the $x$-axis with $x = x_0$, parallel to the $y$-axis $(y = 0)$. Consequently, in contrast to the previously presented types of planes, only orbits with pericenters on the $x$-axis are included and therefore, the value of the dimensionless energy parameter $\widehat{C}$ can be used as an ordinate. In this way, we can monitor how the energy influences the overall orbital structure of the system using a continuous spectrum of energy values rather than few discrete energy levels. In the first row of Fig. \ref{xct} we present the orbital structure of the $(x,\widehat{C})$-plane when $\widehat{C} \in [0.001,0.1]$, while in the second column of the same figure the distribution of the corresponding escape times of orbits is depicted. Unfortunately, Paper I does not contain such an analysis, so we had to compute also the case of the Plummer potential for $M_{\rm cl} = 2.7$ and $c_{\rm P} = 0.38$.

It is seen that in all cases for low values of the energy, above but close enough to the critical escape energy, there is a highly fractal structure between the stability islands. As the value of the energy increases however, the fractality of the planes decreases and several basins of escape start to emerge, mainly at the outer parts of the planes. We can also identify three main stability islands. The one on the left side of the planes correspond to the 1:3 resonant orbits, the middle stability islands contains the 1:1 retrograde orbits, while that on the right side of the plane corresponds to 1:1 prograde orbits. The most weak 1:3 stability region is observed in the Plummer case, while in the case of the Hernquist potential we have the most prominent presence. In all three cases the 1:3 resonance seems to vanish for $\widehat{C} > 0.03$. The stability island of the 1:1 retrograde orbits is more or less the same in all three cases. On the other hand, in the case of the Hernquist potential the stability island of the 1:1 prograde orbits seems to disappear for $\widehat{C} > 0.1$, while at the other two cases is still present. For the Plummer and the isochrone cases we can distinguish the two 3:3 stability regions that embrace the main 1:1 prograde island. It is evident from the results shown in Figs. \ref{xct}(d-f) that the escape times of the orbits are strongly correlated to the escape basins. Moreover, one may conclude that the smallest escape periods correspond to orbits with initial conditions inside the escape basins, while orbits initiated in the fractal regions of the planes or near the boundaries of stability islands have the highest escape rates. In all three cases the escape times of orbits are significantly reduced with increasing energy. Combining all the numerical outcomes presented in Figs. \ref{xct}(d-f) we may say that the key factor that determines and controls the escape times of the orbits is the value of the orbital energy (the higher the energy level the shorter the escape rates), while the fractality of the basin boundaries varies strongly both as a function of the energy and of the spatial variable.

It would be very illuminating to monitor the evolution of the percentages of the different types of orbits on the $(x,\widehat{C})$ planes as a function of the dimensionless energy parameter. For this purpose we introduce Fig. \ref{percs3}(a-c) which contains the results for all three cases. We see that for the Plummer and the isochrone cases the percentage of bounded regular orbits starts at about 55\% and gradually reduces with increasing energy, while in the case of the Hernquist potential the rate of non-escaping regular orbits is about 58\%. In the Plummer case bounded motion is always the most populated type of motion, while in the other two cases escaping orbits prevail at high enough values of the energy. In the interval $0.001 < \widehat{C} < 0.02$ the percentages of escaping orbits seem to fluctuate around 22\% in all three cases, while for larger energy levels the rates start to diverge and orbits escaping through exit channel 2 dominate. It should be emphasized that the greater divergence is observed in the Hernquist case, where escapers form $L_2$ occupy about 45\% of the $(x,\widehat{C})$ plane when $\widehat{C} = 0.1$.

The evolution of the average value of the escape time $< t_{\rm esc} >$ of orbits as a function of the dimensionless energy parameter is given in Fig. \ref{stats}(a) for the $(x,\widehat{C})$ plane. It is seen, that the profiles of the Plummer and the isochrone potential almost coincide, while that of the Hernquist potential has always lower values. In particular, for the Plummer and the isochrone cases and for low values of energy, just above the escape value, the average escape time of orbits is about 800 time units (only 350 for the Hernquist case), however it reduces rapidly tending asymptotically to zero which refers to orbits that escape almost immediately from the system. We feel it is important to justify this behaviour of the escape time. As the value of the Jacobi constant increases the two escape channels (which are of course symmetrical) become more and more wide and therefore, orbits need less and less time in order to find one of the openings in the open equipotential surface and eventually escape from the system. This geometrical feature explains why for low values of energy orbits consume large time periods wandering inside the open equipotential surface until they eventually locate one of the two exits and escape to infinity. Finally Fig. \ref{stats}(b) shows the evolution of the percentage of the total area $(A)$ on the $(x,\widehat{C})$ plane corresponding to basins of escape, as a function of the dimensionless energy parameter. It is seen that for low values of the energy in all three cases the planes are highly fractal ($A$ is almost zero). However, as we proceed to higher energy levels the degree of fractalization reduces and the area corresponding to basins of escape start to grow rapidly. In the interval $0.001 < \widehat{C} < 0.03$ A\% is common in all three cases, while only for $\widehat{C} > 0.03$ the profile of the Hernquist potential start to diverge. Eventually, at very high energy levels $(\widehat{C} = 0.1)$ the fractal domains are significantly confined and therefore the well formed basins of escape occupy about 30\% of the entire plane in the Plummer and the isochrone case and about 44\% in the Hernquist case. Thus we may say that in the Hernquist model orbits escape more quickly and basins of escape are more extended that in the other two models. We assume that the deviations shown in Figs. \ref{stats}(a-b) of the Hernquist potential with respect to the other two (Plummer and isochrone potentials) are mainly due to two reasons which are the only obvious differences between these three potentials: (i) the presence of the central cusp and (ii) the finite force near the center of the star cluster.

\section{Discussion and conclusions}
\label{disc}

In this work we tried to compare the orbital dynamics in star cluster models using three similar potentials for describing the properties of the spherically symmetric star cluster. The work was initiated in Paper I where a Plummer potential was applied, while in the present work we investigated the escape dynamics in the cases of the isochrone and the Hernquist potential. As in Paper I, we restricted our exploration in the two-dimensional model thus using for all orbits $z = \dot{z} = 0$. We managed to distinguish between ordered/chaotic and trapped/escaping orbits and we also located the basins of escape leading to different exit channels, finding correlations with the corresponding escape times of the 2D orbits.

We defined for several values of the Jacobi integral, dense uniform grids of $1024 \times 1024$ initial conditions $(x_0,y_0)$ and $(x_0, \dot{x_0})$ regularly distributed in the area allowed by the energy level on the configuration and phase space, respectively and then we identified regions of order/chaos and bound/escape. For the numerical integration of the orbits in each type of grid, we needed about between 7 hours and 11 days of CPU time on a Pentium Dual-Core 2.2 GHz PC, depending on the escape rates of orbits in each case. For each initial condition, the maximum time of the numerical integration was set to be equal to $10^4$ time units however, when a particle escaped the numerical integration was effectively ended and proceeded to the next available initial condition.

The present paper provides quantitative information regarding the escape dynamics in star cluster models. The main numerical results of our research can be summarized as follows:
\begin{enumerate}
 \item In the Plummer case two types of secondary resonances (the 3:3 and the 4:4 resonant families) surround the main 1:1 stability islands. In the isochrone case only the 3:3 resonant family is present, while in the Hernquist models both secondary resonant families are absent.
 \item The 1:3 resonant family is common in all three cases however, in the Hernquist potential the areas on the configuration and on the phase space covered by 1:3 resonant orbits are much more extended with respect to the other two cases.
 \item At relatively high values of the energy the stability islands corresponding to the 1:3 resonant orbits disappear in all three cases. In addition, the stability island of the 1:1 prograde orbits is confined (isochrone case) or disappear (Hernquist case).
 \item It was observed that the structure of the several escape basins (thin elongated bands or well formed broad regions) is very similar in all three cases, while all differences regarding the particular shapes are very minor.
 \item Our calculations revealed, that the escape times of orbits are directly linked to the basins of escape. In particular, inside the basins of escape as well as relatively away from the fractal domains, the shortest escape rates of the orbits had been measured. On the other hand, the longest escape periods correspond to initial conditions of orbits either near the boundaries between the escape basins or in the vicinity of the stability islands.
 \item In the Hernquist case we identified a non zero amount of trapped chaotic orbits which are mainly located inside the 1:3 stability islands. Additional numerical calculations revealed that these orbits do eventually escape to infinity but only after vast time intervals with no physical meaning.
 \item In the case of the Hernquist models we measured the smallest average escape times in all tested energy levels, while in the cases of the Plummer and the isochrone potentials the average escape rates of orbits are larger. Furthermore, the area of the basins of escape is larger in the Hernquist models. We assume that this divergence is due to the presence of the central cusp in the Hernquist potential.
\end{enumerate}

Judging by the detailed and novel outcomes we may say that our task has been successfully completed. We hope that the present comparison analysis and the corresponding numerical results to be useful in the active field of the dissolution of tidally limited star clusters.

\section*{Acknowledgments}

I would like to express my warmest thanks to Dr. Andreas Ernst and Prof. Douglas C. Heggie for all the illuminating and inspiring discussions during this research. My thanks also go to the anonymous referee for the careful reading of the manuscript and for all the apt suggestions and comments which allowed us to improve both the quality and the clarity of the paper.

\bsp
\label{lastpage}

\end{document}